\documentclass[12pt,a4paper]{article}

\usepackage{amsthm,amsmath,natbib,amssymb,amsfonts,bm}

\usepackage{graphicx}
\usepackage{placeins}
\usepackage{subcaption}
\usepackage{setspace}
\usepackage{float}
\usepackage{mathtools}
\usepackage{pmat}

\graphicspath{{Figures/}}

\usepackage{sectsty} % RM added command to change section heading size - 23/8/17
\sectionfont{\fontsize{12}{16}\selectfont} % RM added command to change section heading size - 23/8/17
\subsectionfont{\fontsize{12}{16}\selectfont} % RM added command to change subsection heading size - 24/8/17

\newtheorem{theorem}{Theorem}

\newtheorem*{theorem*}{Theorem}

\theoremstyle{definition}

\theoremstyle{remark}

\newcommand{\bs}{\boldsymbol}

\DeclareMathAlphabet{\mathpzc}{OT1}{pzc}{m}{it}

\newcommand{\bX}{\boldsymbol{X}}

\newcommand{\bzero}{\boldsymbol{0}}

\newcommand{\bbeta}{\bs{\beta}}

\newcommand{\var}{\mbox{$\textrm{\textup{var}}$}}
\newcommand{\cov}{\mbox{$\textrm{\textup{cov}}$}}
\newcommand{\corr}{\mbox{$\textrm{\textup{corr}}$}}

\newcommand{\blambda}{\boldsymbol{\lambda}}
\newcommand{\by}{\mbox{$\boldsymbol{y}$}}
\newcommand{\bvarepsilon}{\boldsymbol{\varepsilon}}
\newcommand{\CP}{\rm CP}
\newcommand{\SEL}{\rm SEL}

\begin{document}

\baselineskip=21pt

\begin{center}
%\noindent
{\bf \large On confidence intervals centered on bootstrap smoothed estimators}
\end{center}

%\bigskip

%\bigskip

\begin{center}
{\bf Paul Kabaila$^*$ and Christeen Wijethunga}
\end{center}

%\medskip

\begin{center}
%{\large
{\sl Department of Mathematics and Statistics, La Trobe University, Australia}
%}
\end{center}

%\vspace{1cm}
%\medskip

% Maximum of 200 words
\noindent \textbf{ABSTRACT}

%\bigskip

% \medskip

\noindent 
Bootstrap smoothed (bagged) estimators have been proposed as an improvement on estimators found after preliminary data-based model selection. 
Efron, 2014, derived a widely applicable formula for a delta method approximation to the standard deviation of the bootstrap smoothed estimator. He also considered a confidence interval centered on the bootstrap smoothed estimator, with width proportional to the estimate of this standard deviation. 
Kabaila and Wijethunga, 2019, assessed the performance of this confidence interval in the scenario of two nested linear regression models, the full model and a simpler model, for the case of known error variance and preliminary model selection using a hypothesis test. They found that the performance of this confidence interval was not substantially better than the usual confidence interval based on the full model, with the same minimum coverage.
We extend this assessment to the case of unknown error variance
by deriving a computationally convenient exact formula for the ideal (i.e. in the limit as the number of bootstrap replications diverges to infinity) delta method approximation to the standard deviation of the bootstrap smoothed estimator.
Our results show that, unlike the known error variance case, there are circumstances in which this confidence interval has attractive properties.

\medskip


\medskip

\noindent \textbf{1. Introduction}

\smallskip

In applied statistics there is usually some uncertainty as to which explanatory 
variables should be included in the model. The first attempt to deal with this
`model uncertainty' was to use preliminary data-based model selection employing
either hypothesis tests or minimizing a criterion such as the Akaike Information Criterion
(\citeauthor{Akaike1974}, \citeyear{Akaike1974}).
This model selection was followed by the statistical inference of interest, based on the assumption that the selected model had been given to us \textit{a priori},
as the true model. This assumption is false and typically leads to incorrect and misleading inference (see e.g. \citeauthor{Kabaila2009},
\citeyear{Kabaila2009} and \citeauthor{LeebPotscher2005}, \citeyear{LeebPotscher2005}).

Bootstrap smoothed (or bagged; \citeauthor{Breiman1996},
\citeyear{Breiman1996})  estimators have been proposed as an
improvement on estimators found after preliminary data-based model selection (post-model-selection estimators). Bootstrap smoothed estimators are 
smoothed versions of the post-model-selection estimator. 
The key result of \cite{Efron2014} is a formula for a delta method approximation,
${\bf sd}_{\rm delta}$, to the standard deviation of the bootstrap smoothed estimator. 
This formula is valid for any exponential family of models and has the attractive feature that it simply re-uses
the parametric bootstrap replications that were employed to find this estimator.
It also has the attractive feature that it is applicable in the context of complicated data-based model selection.
\cite{KabailaWijethunga2019} consider a confidence interval (CI) centered on the bootstrap
smoothed estimator, with nominal coverage $1 - \alpha$, and half-width equal to the $1 - \alpha/2$ quantile of the standard normal distribution multiplied by the estimate of  ${\bf sd}_{\rm delta}\,$.
We call this interval the ${\bf sd}_{\rm delta}\,{\bf interval}$.

This CI has similarities with the frequentist model averaged CIs proposed
by \cite{BucklandEtAl1997}, \cite{FletcherTurek2011} and \cite{TurekFletcher2012}.
All of these CIs need to have their performances, in terms of coverage probability and expected length, carefully assessed before they can be recommended for general use by applied statisticians. We believe that such assessments are best carried out
through a sequence of increasingly complicated `test scenarios'.

The simplest test scenario consists of 
two nested linear regression models, where the simpler model is given by a specified linear combination of the regression parameters being set to zero. In this test scenario, the scalar parameter of interest is a distinct linear combination of the regression parameters and we assume independent and identically distributed normal errors, with error variance assumed known.
\cite{KabailaWijethunga2019} provide a detailed assessment of the performance of the ${\bf sd}_{\rm delta}\,{\bf interval}$ in this test scenario if the
simpler model is selected when a preliminary hypothesis test accepts the null hypothesis that this simpler model is correct. 
They found that, while this CI performed much better than the post-model-selection confidence interval in terms of minimum coverage 
probability, its performance in terms of expected length was not substantially better than the usual CI based on the full model, with the same minimum coverage.

The next simplest test scenario is the same, but with 
unknown error variance. \cite{KabailaWelshAbeysekera2016} and \cite{KabailaWelshMainzer2017} used this test scenario to provide a detailed assessment of the performance of the CIs proposed by \cite{FletcherTurek2011} and \cite{TurekFletcher2012}.
%\cite{HjortClaes2003} use the test scenario of two nested general regression models,
%with the simpler model specified by a single (possibly nonlinear) constraint on the parameters of the model, to provide a large sample assessment of the CI proposed by \cite{BuckEtAl1997}.
Our aim is to extend the assessment 
made by \cite{KabailaWijethunga2019} of the performance of the 
${\bf sd}_{\rm delta}\,{\bf interval}$ to this test scenario.

We apply Theorem 2 of \cite{Efron2014}
to derive a computationally convenient exact formula for the ideal (i.e. in the limit as the number of bootstrap replications diverges to infinity) delta method approximation to the standard deviation of the bootstrap smoothed estimator.
An outline of this derivation, which is quite complicated, is provided in Appendix A.1.
Our computed results show that, unlike the case that the error variance is assumed known, there are circumstances 
%(for the 
%test scenario under consideration)
%case of unknown error variance) 
in which the expected length properties of the
${\bf sd}_{\rm delta}\,{\bf interval}$ are quite attractive.

\medskip

\noindent \textbf{2. The two nested regression models and the post-model-selection estimator}

\smallskip

We consider two nested linear regression models: the full model
${\cal M}_2$ and the simpler model ${\cal M}_1$.
Suppose that the full model ${\cal M}_2$ is given by 
\begin{equation*}
%\label{SimplerScenario}
%\label{m2}
\by = \bX \bbeta + \bvarepsilon
\end{equation*}
%
%%
%\begin{equation*}
%\widetilde {\by} = \widetilde {\bX} \bvarphi + \bepsilon
%\end{equation*}
%%
where $\by$ is a random $n$-vector of responses, $\bX$ is a known $n \times p$ matrix with linearly independent columns 
($p < n$), $\bbeta$ is an unknown $p$-vector of parameters and $\bvarepsilon \sim N(\bzero, \sigma^2 \boldsymbol{I})$, with $\sigma^2$ an unknown positive parameter.
Suppose that 
$\bbeta = [\theta, \tau, \boldsymbol\lambda^\top ]^\top$, where $\theta$ is the scalar parameter of interest, 
$\tau$ is a scalar parameter used in specifying the model ${\cal M}_1$ and $\blambda$ is a ($p-2$)-dimensional parameter vector. The model ${\cal M}_1$ is ${\cal M}_2$ with $\tau = 0$. As shown in Appendix A of  \cite{KabailaWijethunga2019}, this scenario can be obtained by a change of parametrization from a more
general scenario. Let $m=n-p$.

Let $\widehat{\bbeta}$ denote the least squares estimator of $\bbeta$, so that
$\widehat{\bbeta} = (\bX^\top \bX)^{-1} \bX^\top \by$,
and $\widehat{\sigma}^2 = (\by - \bX \widehat\bbeta)^{\top} (\by - \bX \widehat\bbeta)/m$.
Also let $\widehat{\theta}$ and $\widehat{\tau}$ denote the first and second components of
$\widehat{\bbeta}$, respectively.
Now
let $v_\theta = \var(\widehat{\theta})/ \sigma^2$, $v_\tau = \var(\widehat{\tau}) /\sigma^2$ and
$\rho=\corr(\widehat{\theta}, \widehat\tau) = v_{\theta\tau}/(v_\theta v_\tau)^{1/2}$, where $v_{\theta\tau} = \cov(\widehat{\theta}, \widehat{\tau})/\sigma^2$.
Note that $v_{\theta}$, $v_\tau$,  $v_{\theta\tau}$ and $\rho$ are known.
Let $\gamma = \tau / \big(\sigma v_\tau^{1/2} \big)$, which is an unknown parameter, and
$\widehat{\gamma} = \widehat{\tau}/(\widehat\sigma{v_\tau}^{1/2})$.

Suppose that we carry out a preliminary test of the null hypothesis $\tau = 0$ against the alternative hypothesis
$\tau \neq 0$
and that we choose the model ${\cal M}_1$ if this null hypothesis is accepted;
otherwise we choose the model ${\cal M}_2$.
%The test statistic is $|\widehat{\gamma}|$, which has the same distribution as $|T|$, for $|T| \sim t_m$, under the null hypothesis. 
Let $t_m(a)$ be defined by $P(T \le t_m(a)) = 1 - a/2$ for $T \sim t_m$.
Suppose that we accept the null hypothesis when 
$|\widehat{\gamma}| \leq t_m(\widetilde{\alpha})$; otherwise we reject the null hypothesis. The size of this 
preliminary test is $\widetilde{\alpha}$. 
Therefore the post-model-selection estimator of $\theta$ is equal to 
\begin{equation*}
\widehat{\theta}_{\textsc{\tiny PMS}} 
= \begin{cases}
\widehat{\theta} - \displaystyle{\frac{v_{\theta\tau}}{v_\tau}} \widehat{\tau} \qquad \text{\ \ \ \ if } \ |\widehat{\gamma}| \leq t_m(\widetilde{\alpha})
\\
\widehat{\theta} \qquad \text{\ \ \ \ otherwise}.
\end{cases}
\end{equation*}
Henceforth,  suppose that 
$1-\alpha$ 
and $\widetilde{\alpha}$ are given.

\medskip

\noindent \textbf{3. Computationally convenient exact formulas for the ideal bootstrap smoothed estimate and the delta method approximation to its standard deviation}

\smallskip

The parametric bootstrap smoothed estimate of $\theta$ is obtained as follows.
Note that $\widehat{\bbeta} \sim N \big(\bbeta, \sigma^2 (\bX^{\top} \bX)^{-1} \big)$
and, independently, $m^{1/2} \widehat{\sigma} / \sigma \sim \chi_m$ (if $Q \sim \chi_m^2$ then $Q^{1/2}$ is said to have a $\chi_m$ distribution).
To make the dependence of $\widehat{\theta}_{\textsc{\tiny PMS}}$ on 
$(\widehat{\bbeta}, \widehat{\sigma})$ explicit, write 
$\widehat{\theta}_{\textsc{\tiny PMS}} = g(\widehat{\bbeta}, \widehat{\sigma})$.
For the estimate $(\widehat{\bbeta}, \widehat{\sigma})$  treated as the true parameter value, suppose that 
$\widehat{\bbeta}^* \sim N \big(\widehat{\bbeta}, \widehat{\sigma}^2 (\bX^{\top} \bX)^{-1} \big)$
and, independently, $m^{1/2} \widehat{\sigma}^* / \widehat{\sigma} \sim \chi_m$. 
A parametric bootstrap sample of size $B$ consists of independent observations
$\big(\widehat{\bbeta}_1^*,\widehat{\sigma}_1^*\big),
\big(\widehat{\bbeta}_2^*,\widehat{\sigma}_2^*\big), \dots, \big(\widehat{\bbeta}_B^*,\widehat{\sigma}_B^*\big),$ of the random vector
$\big(\widehat{\bbeta}^*,\widehat{\sigma}^*\big)$.
The parametric smoothed estimate of $\theta$ is defined to be
\begin{equation*}
\frac{1}{B} \sum_{i=1}^B g\big(\widehat{\bbeta}_i^*,\widehat{\sigma}_i^*\big).
\end{equation*}
The limit as the number
of boostrap replications $B \rightarrow \infty$
of this quantity is called by \cite{Efron2014}  the ideal bootstrap smoothed estimate of $\theta$. We denote this ideal boostrap smoothed estimate by $\widetilde{\theta}$ and observe that it may be obtained as follows. 
Let $E_{\bbeta, \sigma}(\widehat{\theta}_{\textsc{\tiny PMS}})$ denote the expected value of $\widehat{\theta}_{\textsc{\tiny PMS}}$,
for true parameter value $(\bbeta, \sigma)$.
The ideal bootstrap smoothed estimate
$\widetilde{\theta}$ is obtained by first evaluating 
$E_{\bbeta, \sigma}(\widehat{\theta}_{\textsc{\tiny PMS}})$ and then replacing
$(\bbeta, \sigma)$ by $\big(\widehat{\bbeta}, \widehat{\sigma} \big)$.

Let  $W = \widehat{\sigma}/\sigma$ and define $k_m(\gamma)$ to be
\begin{equation}
\label{Defn_km}
\int_{0}^{\infty} \Big( \phi(d_m w + \gamma) - \phi(d_m w - \gamma) + \gamma \big( \Phi(d_m w - \gamma) - \Phi(-d_m w - \gamma) \big) \Big)\, f_W(w)\, dw, 
\end{equation}	
where $\phi$ and $\Phi$ denote the $N(0,1)$ pdf and cdf, respectively, $d_m = t_m(\widetilde{\alpha})$ and $f_W$ denotes the probability density function of $W$. As proved in Appendix B of \cite{KabailaWijethunga2019},
$E_{\bbeta, \sigma}(\widehat{\theta}_{\textsc{\tiny PMS}}) = \theta - \rho \, \sigma \, v_\theta^{1/2} \, k_m(\gamma)$. Therefore
\begin{equation*}
\widetilde{\theta} 
= \widehat{\theta} - \rho \, \widehat{\sigma}\, v_\theta^{1/2} \, k_m(\widehat{\gamma}).
\end{equation*}

An outline of the proof of the following new theorem is given in Appendix A.1.

\begin{theorem}
	\label{TheoremIdealBootstrapSmoothedEstimatorSDdeltaEfron}
	An application of Theorem 2 of \cite{Efron2014} leads to the ideal 
	(i.e. in the limit as the number
	of boostrap replications $B \rightarrow \infty$) delta method approximation 
	to the standard deviation of $\widetilde{\theta}$, denoted by 
	${\bf sd}_{\rm delta}(\gamma, \sigma)$, which is
	$\sigma v_{\theta}^{1/2} \, r_{\, \rm delta}(\gamma)$, where
	\begin{equation}
	\label{Defn_rdelta}
	r_{\, \rm delta}(\gamma) = \Bigg( \frac{\rho^2}{2n} \Big( k_m(\gamma) + h_m(\gamma) - \gamma \, q_m(\gamma) \Big)^2 + 1 - 2\rho^2 q_m(\gamma) + \rho^2 q_m^2(\gamma) \Bigg)^{1/2}.
	\end{equation}
	Here $q_m(\gamma)$ is defined to be
	\begin{equation}
	\label{Defn_qm}
	\int_{0}^{\infty} \big( -d_m w\, \phi(d_m w+\gamma) - d_m w\, \phi(d_m w-\gamma) + \Phi(d_m w-\gamma) - \Phi(-d_m w-\gamma) \big)\, f_W(w)\, dw
	\end{equation}
	and
	\begin{equation}
	\label{Defn_hm}
	h_m(\gamma) 
	= \int_{0}^{\infty} \Big( (d_m w)^2 \phi(d_m w + \gamma) - (d_m w)^2 \phi(d_m w - \gamma) \Big)\, f_W(w)\, dw,
	\end{equation}
	where, as before, $d_m = t_m(\widetilde{\alpha})$.
\end{theorem}

We expect, intuitively, that the results obtained for the case that $\sigma^2$ is
unknown (so that it must be estimated from the data) and $m \rightarrow \infty$ should be the same as for the case that $\sigma^2$ is known. 
Suppose that $p$ is fixed and $n \rightarrow \infty$, so that $m=n-p$ also diverges to $\infty$. As expected, the ideal delta method approximation to the standard deviation of $\widetilde{\theta}$ given by Theorem 1 converges to the 
corresponding quantity
% ideal delta method approximation to the standard deviation of
% $\widetilde{\theta}$ 
given by Theorem 2 of \cite{KabailaWijethunga2019}, which deals with the case that 
$\sigma^2$ is known.

\medskip

\noindent \textbf{4. Computationally convenient exact formula for the coverage probability of the confidence interval centered on the bootstrap smoothed estimator}

\smallskip

Consider the CI for $\theta$ centered on the bootstrap smoothed estimator $\widetilde{\theta}$,  with nominal coverage $1-\alpha$,
\begin{align*}
J_{\rm delta} &= \Big[ \widetilde{\theta} - t_m(\alpha)\, {\bf sd}_{\rm delta}(\widehat\gamma, \widehat\sigma), \, 
\widetilde{\theta} + t_m(\alpha)\, {\bf sd}_{\rm delta}(\widehat\gamma, \widehat\sigma) \Big]\\
&= \Big[ \widetilde{\theta} - t_m(\alpha)\, \widehat\sigma \, v_{\theta}^{1/2} \, r_{\, \rm delta}(\widehat\gamma), \, 
\widetilde{\theta} + t_m(\alpha)\, \widehat\sigma \, v_{\theta}^{1/2} \, r_{\, \rm delta}(\widehat\gamma) \Big],
\end{align*}
which we call the ${\bf sd}_{\rm delta}$ {\bf interval}. 
Note that when $\rho = 0$, this CI is identical to the usual CI, with actual coverage $1-\alpha$, based on the full model ${\cal M}_2$. 
It may be shown that the coverage probability $P(\theta \in J_{\rm delta})$
is a function of $(\gamma, \rho)$. We therefore denote this coverage probability by
${\CP}_{\rm delta}(\gamma, \rho)$. The following theorem is proved in Appendix A.2.
\begin{theorem}
	\label{CPdeltaEfronVarUnknown}
	Let 
	\begin{align}
	\label{Def_ell}
	\ell(h, w, \rho) &= -w\, t_m(\alpha)\, r_{\rm delta}\left(\frac{h}{w} \right) + w\,\rho \, \,  k_m\left(\frac{h}{w} \right) 
	\\
	\label{Def_u}
	u(h, w, \rho) &= w\, t_m(\alpha)\,  r_{\rm delta}\left(\frac{h}{w} \right) + w\,\rho \, \, k_m\left(\frac{h}{w} \right).
	\end{align}
	Then ${\CP}_{\rm delta}(\gamma, \rho)$ is given by
	\begin{equation*}
	\int_{0}^{\infty} \int_{-\infty}^{\infty} \Psi \Big( \ell(y+\gamma, w, \rho), u(y+\gamma, w, \rho); \rho(y), 1-\rho^2 \Big) \phi(y) \, dy \, f_W(w)\, dw,
	\end{equation*}
	where $\Psi\big(\ell, u; \mu, v \big) = P \big(\ell \leq Z \leq u\big)$ for $Z \sim N(\mu, v)$.
	
\end{theorem} 

The expression \eqref{Defn_rdelta} suggests that, for all sufficiently large $n$,
${\CP}_{\rm delta}(\gamma, \rho)$ is determined by $m$, for any given $(\gamma, \rho)$.
Computational results for
$n=25$ (described later in this section) and $n=100$ (not described either here or in the Supporting Material) suggest that, for all $n \ge 25$, 
${\CP}_{\rm delta}(\gamma, \rho)$ is, for practical purposes, determined by $m$, for any given $(\gamma, \rho)$.
It may be shown that ${\CP}_{\rm delta}(\gamma, \rho)$ is (a) an even function of $\gamma$ for each $\rho$ and (b) an even function of $\rho$ for each $\gamma$. 
It follows that,
for given $n$ and $m$,
we are able to encapsulate the coverage probability of 
the
${\bf sd}_{\rm delta}\,{\bf interval}$, for all possible choices of design matrix, parameter of interest $\theta$ and parameter $\tau$ that specifies the simpler model, using only the parameters $|\rho|$ and $|\gamma|$.

\begin{figure}[h]
	\centering
	\includegraphics[width=0.9\textwidth]{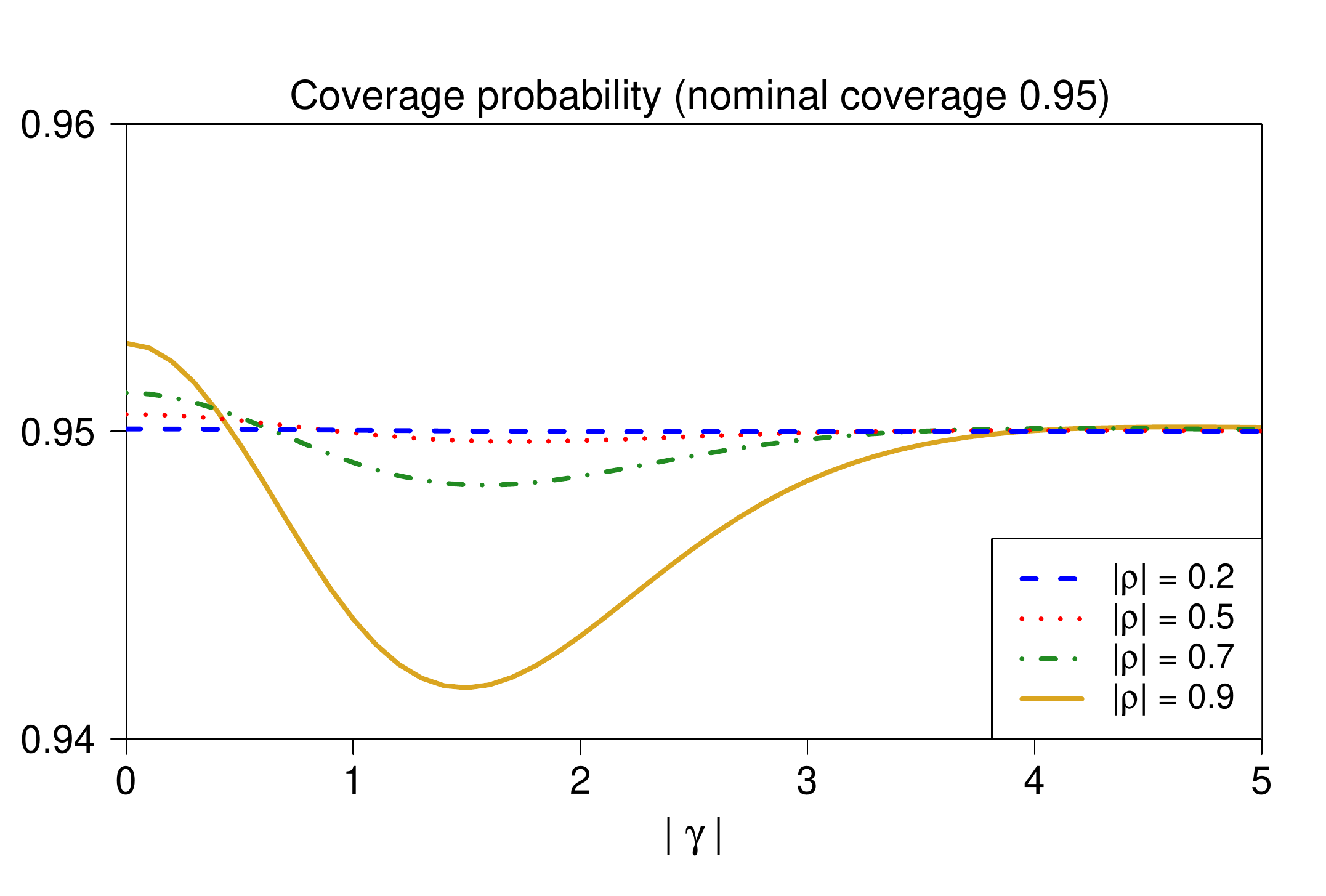}
	\caption{The coverage probability of the ${\bf sd}_{\rm delta}$ {\bf interval},
		which is based on the post-model-selection estimator obtained after a preliminary hypothesis test, with size $0.1$, of the null hypothesis that the simpler model is correct. The nominal coverage is 0.95,
		$n=25$, $m=1$ and $|\rho| = 0.2, 0.5, 0.7$ and 0.9.}
	\label{CP_Efdelta_n_25_m_1}
\end{figure}

Figure \ref{CP_Efdelta_n_25_m_1} is the graph of coverage probability of the confidence interval $J_{\rm delta}$ centered on the bootstrap smoothed estimator, which is based on the post-model-selection estimator obtained after a preliminary hypothesis test, with size $\widetilde{\alpha} = 0.1$, of the null hypothesis that the simpler model is correct. We consider the case 
that the nominal coverage is 0.95,
$n=25$, $m=1$ and $|\rho| = 0.2, 0.5, 0.7$ and 0.9. 
All of the computations reported in this paper were carried out using programs written in
\texttt{R}.
The minimum coverage probability of this CI is a continuous decreasing function of $|\rho|$
which equals the nominal coverage 
%of this CI 
when $\rho =0$. Graphs of the coverage probability of $J_{\rm delta}$
for the same values of nominal coverage, size of the preliminary hypothesis test, $n$ and $|\rho|$ are 
provided 
in the Supporting Material for $m=2, 3$ and 10. 
Further extensive numerical investigations, not reported either here
or in the Supporting Material, show that the 
${\bf sd}_{\rm delta}\,{\bf interval}$ outperforms the post-model-selection CI, with the same nominal coverage and based on the same preliminary test, in terms of coverage probability. 

\medskip

\noindent \textbf{5. Computationally convenient exact formula for the scaled expected length of the confidence interval centered on the bootstrap smoothed estimator}

\smallskip

We define the scaled expected length of $J_{\rm delta}$, with nominal coverage $1-\alpha$, 
to be the expected length of $J_{\rm delta}$ divided by the expected length of the usual CI, based on the full model, with the same coverage
as the minimum coverage probability of $J_{\rm delta}$. Let $c_{\rm min}$ denote this minimum coverage probability. Now let $I(c)$ denote the usual CI for $\theta$, with coverage probability $c$, based on the full model. In other words, 
$I(c) = \Big[ \widehat{\theta} - t_m(1-c)\, \widehat{\sigma}\, v_\theta^{1/2}, \, 
\widehat{\theta} + t_m(1-c)\, \widehat{\sigma}\, v_\theta^{1/2} \Big]$. 
It may be shown that the scaled expected length of $J_{\rm delta}$
is a function of $(\gamma, \rho)$. We therefore denote this scaled expected length by ${\SEL}_{\rm delta}(\gamma, \rho)$.
The following theorem is proved in Appendix A.3.
\begin{theorem}
	\label{SELdeltaEfronVarUnknown}
	Let $c_{\rm min}$ denote the minimum coverage probability of the confidence interval $J_{\rm delta}$, with nominal coverage $1-\alpha$. Then ${\SEL}_{\rm delta}(\gamma, \rho)$ is given by
	\begin{equation*}
	\frac{t_m(\alpha)}{t_m(1-c_{\rm min})}\, \left(\frac{m}{2}\right)^{1/2}\frac{\Gamma(m/2)}{\Gamma((m+1)/2)}\, \int_{0}^{\infty} \int_{-\infty}^{\infty} w\,\, r_{\rm delta}\left(\frac{y+\gamma}{w} \right)\, \phi(y)\, dy\, f_W(w)\, dw.
	\end{equation*}
	
\end{theorem}

The expression \eqref{Defn_rdelta} suggests that, for all sufficiently large $n$,
${\SEL}_{\rm delta}(\gamma, \rho)$ is determined by $m$, for any given $(\gamma, \rho)$.
Computational results for
$n=25$ (described later in this section) and $n=100$ (not described either here or in the Supporting Material) suggest that, for all $n \ge 25$, 
${\SEL}_{\rm delta}(\gamma, \rho)$ is, for practical purposes, 
determined by $m$, for any given $(\gamma, \rho)$.
It may be shown that ${\SEL}_{\rm delta}(\gamma, \rho)$ is (a) an even function of $\gamma$ for each $\rho$ and (b) an even function of $\rho$ for each $\gamma$. 
It follows that,
for given $n$ and $m$, we are able to encapsulate the scaled expected length of 
the 
${\bf sd}_{\rm delta}\,{\bf interval}$, for all possible choices of design matrix, parameter of interest $\theta$ and parameter $\tau$ that specifies the simpler model, 
using only the 
parameters $|\rho|$
and $|\gamma|$.

The bootstrap smoothed estimator is obtained by smoothing the post-model-selection estimator that results from a preliminary test of the null hypothesis that the simpler model is correct i.e. that $\gamma = 0$.
This post-model-selection estimator is usually motivated by a desire for 
good performance when the simpler model is correct. Therefore, ideally, the 
${\bf sd}_{\rm delta}\,{\bf interval}$ should have a scaled expected length that is substantially less than 1 when $\gamma = 0$. In addition, ideally, this confidence interval should have a scaled expected length that (a) has maximum value that is not too much larger than 1 and (b) approaches 1 as $|\gamma|$ approaches infinity. 

\begin{figure}[h]
	\centering
	\includegraphics[width=0.9\textwidth]{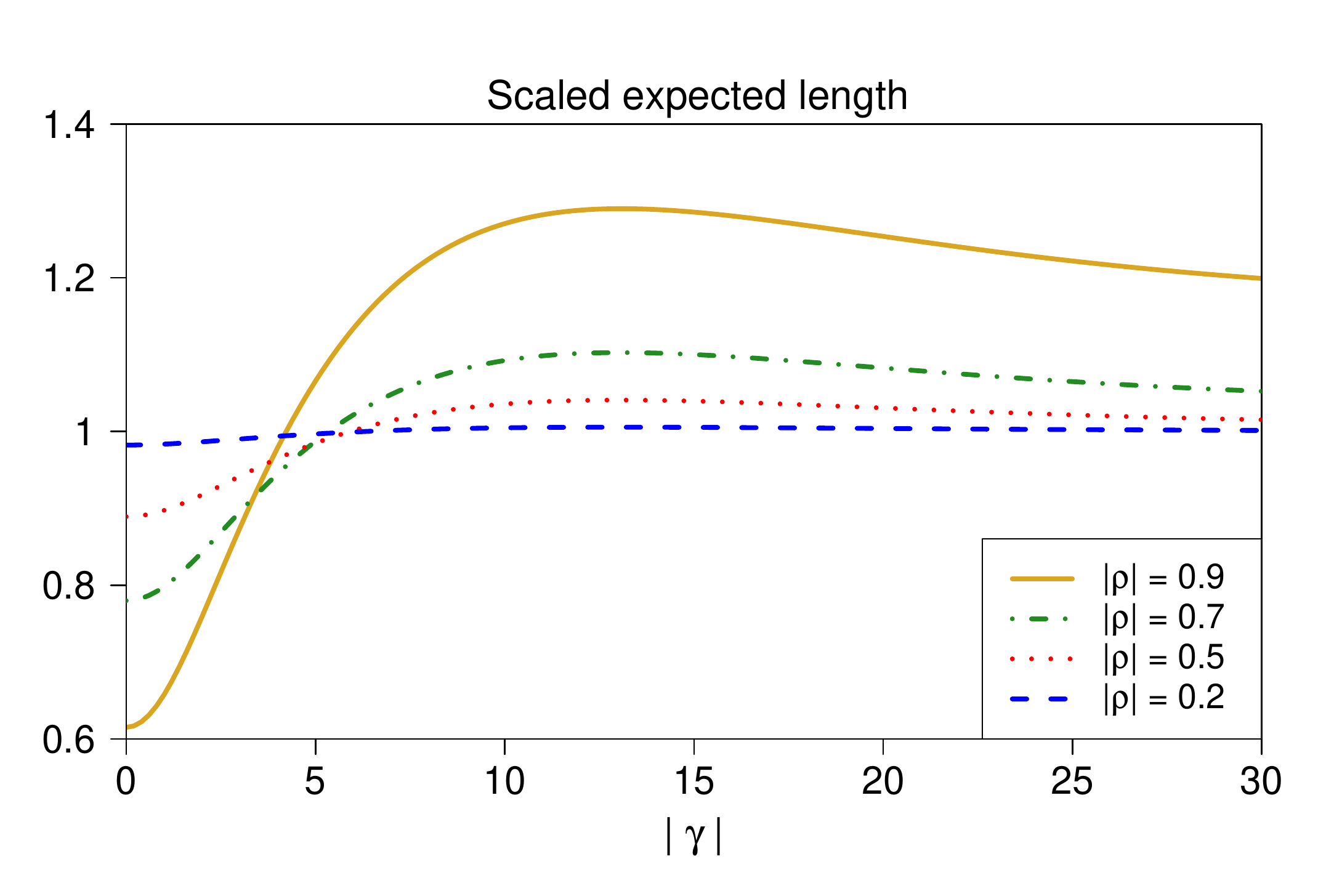}
	\caption{The scaled expected length of the ${\bf sd}_{\rm delta}$ {\bf interval},
		which is based on the post-model-selection estimator obtained after a preliminary hypothesis test, with size $0.1$, of the null hypothesis that the simpler model is correct. The nominal coverage is 0.95,
		$n=25$, $m=1$ and $|\rho| = 0.2, 0.5, 0.7$ and 0.9.}
	\label{SEL_Efdelta_n_25_m_1}
\end{figure}

Figure \ref{SEL_Efdelta_n_25_m_1} is the graph of scaled expected length of the confidence interval centered on the bootstrap smoothed estimator, which is based on the post-model-selection estimator obtained after a preliminary hypothesis test, with size $\widetilde{\alpha} = 0.1$, of the null hypothesis that the simpler model is correct. We consider the case 
that the nominal coverage is 0.95,
$n=25$, $m=1$ and $|\rho| = 0.2, 0.5, 0.7$ and 0.9. 
For $|\rho| = 0.5, 0.7$ and 0.9, the scaled expected length is substantially less than 1 when $\gamma = 0$. In addition, the scaled expected length (a) has maximum value that is not too much larger than 1 and (b) approaches 1 as $|\gamma|$ approaches infinity. This shows that for $m=1$ and $|\rho| \ge 0.5$ 
the scaled expected length of ${\bf sd}_{\rm delta}$ {\bf interval} has the desired properties. This finding is similar to that reported in \cite{KabailaGiri2013} 
concerning the performance of the CIs constructed by 
\cite{KabailaGiri2009a} to have the desired coverage probability and these desired scaled expected length properties. Namely, the performance of this CI improves as $|\rho|$ increases and $m$ decreases.

By contrast, for the case that $\sigma^2$ is assumed known, examined by \cite{KabailaWijethunga2019},
the 
scaled expected length of the CI centered on the bootstrap smoothed estimator  (a)  
is either greater than 1 or only
slightly less than 1 at $\gamma = 0$ and (b) has maximum value that is an increasing function of $|\rho|$ that can be much larger than 1 for large $|\rho|$.
As noted earlier, we expect that as $m$ increases (which implies that $n$ also increases), the results obtained in the present paper will approach the corresponding results obtained by \cite{KabailaWijethunga2019}. Therefore we expect that as $m$ increases the ${\bf sd}_{\rm delta}$ {\bf interval} will get further and further away from possessing the desired scaled expected length properties.    
This is confirmed by the graphs of the scaled expected length of $J_{\rm delta}$
for nominal coverage 0.95, size $\widetilde{\alpha} = 0.1$ of the preliminary hypothesis test, $n=25$ and $|\rho| \in \{0.2, 0.5, 0.7, 0.9\}$ that are 
provided 
in the Supporting Material for $m=2, 3$ and 10.

\medskip

\noindent \textbf{6. Discussion}

\smallskip

For the test scenario of two nested linear regression models and error variance assumed known, \cite{KabailaWijethunga2019} found that the  ${\bf sd}_{\rm delta}$ {\bf interval} does not perform any better in terms of expected length than the usual confidence interval, with the same minimum coverage probability and based on the full model. Intuitively, the case that the error variance is assumed to be known corresponds to the case that the error variance is unknown (so that it must be estimated) and the number of degrees of freedom $m$ for the estimation of the error variance is large. 

In the present paper, we deal with the case 
that the error variance is unknown. We find that, for small $m$ and large magnitude of correlation between the least squares estimators of the parameter of interest and the parameter that is set to zero to specify the simpler model, the expected length of the  ${\bf sd}_{\rm delta}$ {\bf interval} possesses some attractive features. 

\medskip

\noindent \textbf{Acknowledgement}

%\noindent 
This work was supported by an Australian Government Research Training Program Scholarship.

%\noindent \textbf{Appendix}

\baselineskip=21pt

\medskip

\noindent \textbf{Appendix}

\smallskip

Let $\widetilde{\gamma}=\widehat{\tau}/(\sigma{v_\tau}^{1/2})$, so that 
$\widehat{\gamma} = \widetilde{\gamma}/W$, where
$W = \widehat{\sigma} / \sigma$. Note that $(\widehat{\theta}, \widetilde{\gamma})$ and $W$ are independent and $W$ has the same distribution as $(Q/m)^{1/2}$ where $Q \sim \chi^2_m$.
To find convenient formulas for expectations and probabilities of interest, we will express all
quantities of interest in terms of $W$ and the random vector 
$\big(\widehat{\theta}, \widetilde{\gamma} \big)$, which has
a bivariate normal distribution with mean $(\theta, \gamma)$ and known covariance matrix with diagonal elements 1 and off-diagonal elements $\rho$.

\medskip

\noindent \textbf{A1. Outline of the Proof of Theorem \ref{TheoremIdealBootstrapSmoothedEstimatorSDdeltaEfron}}

\smallskip

\noindent For the sake of brevity, we present only an outline of the proof of Theorem 1.
By (1.6) of \cite{BNandCox1994}, the pdf
of $\bm{y}$ can be expressed in the exponential family form 
$h(\bm{y}) \exp \big( \widehat{\bm{s}}^\top \bm{\eta}  - \psi(\bm{\eta}) \big)$, where $\widehat{\bm{s}}$ is a sufficient statistic and $\bm{\eta}$ is the unknown parameter vector, with
\begin{align*}
\widehat{\bm{s}} = \left[ \begin{array}{c}
\bm{y}^\top \bm{y} \\
\widehat{\bbeta}
\end{array}\right], \quad
\bm{\eta} = \left[ \begin{array}{c}
-1/(2\sigma^2)\\
\bm{X}^\top \bm{X} \bm{\beta}/\sigma^2
\end{array} \right] \quad \text{and} \quad
\psi(\bm{\eta}) = \frac{\bbeta^\top \bm{X}^\top \bm{X} \bbeta}{2\sigma^2} + \frac{n}{2} \log(\sigma^2).
\end{align*} 

For any two random vectors $\bm{u}$ and $\bm{v}$, define 
$\cov(\bm{u}, \bm{v}) 
= E \big((\bm{u} - E(\bm{u})) (\bm{v} - E(\bm{v}))^{\top}\big)$.
By Theorem 2 of \cite{Efron2014}, the ideal delta method approximation 
to the standard deviation of $\widetilde{\theta}$, which we denote by 
${\bf sd}_{\rm delta}$, is given by 
\begin{equation}
\label{DeltaMethodSD}
{\bf sd}_{\rm delta} = \Big( \big(\cov_*(\bm{\eta})\big)^\top \, \big(V(\bm{\eta}) \big)^{-1} \, \cov_*(\bm{\eta}) \Big)^{1/2},
\end{equation} 
where $\cov_*(\bm{\eta}) = \cov \big(\widehat{\bm{s}}, \, \widehat{\theta}_{\textsc{\tiny PMS}} \big)$
and
\begin{align*}
V(\bm{\eta}) &= \cov (\widehat{\bm{s}}) = \cov \left[ \begin{array}{c}
\bm{y}^\top \bm{y} \\
\widehat{\bbeta}
\end{array} \right] 
= \begin{pmat}[{|}]
\var\big(\bm{y}^\top \bm{y} \big) & \cov \big( \bm{y}^\top \bm{y}, \widehat{\bbeta} \big) \cr\-
\cov \big( \widehat{\bbeta}, \bm{y}^\top \bm{y} \big) & \cov\big(\widehat{\bbeta}\big) \cr
\end{pmat}.
\end{align*}
Now $\bm{y}^\top \bm{y} - E(\bm{y}^\top \bm{y}) = 2\bbeta^\top \bm{X}^\top \bm{\varepsilon} + \bm{\varepsilon}^\top \bm{\varepsilon} - n\sigma^2$. Thus
\begin{align*}
\var(\bm{y}^\top \bm{y}) 
&= E \left( \Big( \bm{q}^\top \bvarepsilon + \sum_{i=1}^{n} (\varepsilon_i^2 - \sigma^2) \Big)^2 \right), 
\quad \text{where} \quad \bm{q}^\top = 2\bm{\beta}^\top \bm{X}^\top,
\\
&= 4\sigma^2 \bbeta^\top \bm{X}^\top \bm{X} \bbeta + 2n\sigma^4.
\end{align*}
Also, $\cov\big( \bm{y}^\top\bm{y}, \widehat\bbeta \big)= 2\sigma^2 \bbeta^\top$. Thus
\begin{align*}
%\label{V_eta}
V(\bm{\eta}) 
= \sigma^2 \begin{pmat}[{|}]
4 \bbeta^\top \bm{X}^\top \bm{X} \bbeta + 2n\sigma^2  & 2 \bbeta^\top \cr\-
2 \bbeta &  \big( \bm{X}^\top \bm{X}\big)^{-1} \cr
\end{pmat}.
\end{align*}
Hence
%
%\begin{equation}
%\label{V_eta_inverse}
%\big(V(\bm{\eta})\big)^{-1} =  \frac{1}{\sigma^2} \left[ \begin{array}{c c}
%\displaystyle \frac{1}{2n\sigma^2} & \quad -\displaystyle \frac{1}{n\sigma^2}  \bbeta^\top \bX^\top \bX\\\\
%- \displaystyle\frac{1}{n\sigma^2} \bX^\top \bX \bbeta & \quad \Big( \boldsymbol{I} + \displaystyle\frac{2}{n\sigma^2}  \bX^\top \bX \bbeta  \bbeta^\top\Big) \bX^\top \bX
%\end{array} \right].
%\end{equation}
%
\begin{equation}
\label{V_eta_inverse}
\big(V(\bm{\eta})\big)^{-1} 
= \frac{1}{\sigma^2} \begin{pmat}[{|}]
\displaystyle \frac{1}{2n\sigma^2}  & -\displaystyle \frac{1}{n\sigma^2}  \bbeta^\top \bX^\top \bX \cr\-
- \displaystyle\frac{1}{n\sigma^2} \bX^\top \bX \bbeta &  \Big( \boldsymbol{I} + \displaystyle\frac{2}{n\sigma^2}  \bX^\top \bX \bbeta  \bbeta^\top\Big) \bX^\top \bX \cr
\end{pmat}.
\end{equation}

\medskip

\noindent Let $\bm{s} = E(\bm{\widehat{s}})$ and observe that $\cov_*(\bm{\eta})$
is equal to 
\begin{align*}
&E \Big( \big( \widehat{\bm{s}} - \bm{s} \big) \big( \widehat{\theta}_{\textsc{\tiny PMS}} - \theta \big) \Big)\\
&= E \Big( \big( \widehat{\bm{s}} - \bm{s} \big) \big( \widehat{\theta} - \theta \big) \Big) - \rho\, \sigma\, v_\theta^{1/2} \int_{0}^{\infty} \int_{-d_m w}^{d_m w} z \, E\big( \widehat{\bm{s}} - \bm{s} \big| \widetilde{\gamma}=z \big) \phi(z-\gamma)\, dz \, f_W(w)\, dw
\end{align*}
where $d_m = t_m(\widetilde\alpha)$. It may be shown that
\begin{align*}
E \Big( \big(\widehat{\bm{s}} - \bm{s}\big) \big( \widehat{\theta} - \theta \big) \Big) 
%&=
%E \left( \left[ \begin{array}{c}
%\by^\top \by - E\big( \by^\top \by \big)\\
%\widehat{\bbeta} - \bbeta
%\end{array}\right] \Big( \widehat{\theta} - \theta \Big) \right)\\
%&=  \left[ \begin{array}{c}
%E \Big( \big( 2\bbeta^\top \bX^\top \bvarepsilon + \bvarepsilon^\top \bvarepsilon - n\sigma^2 \big) \big(  \widehat{\theta} - \theta \big) \Big) \\
%E \Big( \big(\widehat{\bbeta}-\bbeta\big) \big( \widehat{\theta} - \theta \big) \Big)
%\end{array} \right]\\
%&= \left[ \begin{array}{c}
%2\, \sigma^2 \, \theta \\
%\sigma^2 v_\theta^{1/2} \left[ \begin{array}{c}
%v_\theta^{1/2}\\
%\rho\, v_\tau^{1/2}\\
%\bzero
%\end{array} \right] 
%\end{array} \right]\\
%&
= \sigma^2 \left[ \begin{array}{c}
2\, \theta\\
v_\theta\\
\rho\, v_\theta^{1/2}\, v_\tau^{1/2}\\
\bzero
\end{array} \right]   
\end{align*}
and
\begin{align*}
E \big( \widehat{\bm{s}} - \bm{s} \big| \widetilde{\gamma}=z \big) 
%&= E \left( \left[ \begin{array}{c}
%\by^\top \by - E\big( \by^\top \by \big)\\
%\widehat{\bbeta} - \bbeta
%\end{array}\right] \, \vast| \, \widetilde{\gamma}=z \right)\\
%&= \left[ \begin{array}{c}
%E \Big( \by^\top \by - E\big( \by^\top \by \big) \, \Big| \, \widetilde{\gamma}=z \Big)\\
%E \Big( \widehat{\bbeta} - \bbeta \, \Big| \, \widetilde{\gamma}=z \Big)
%\end{array} \right] \\
&= \left[ \begin{array}{c}
E \Big( \by^\top \by - E\big( \by^\top \by \big) \, \Big| \, \widetilde{\gamma}=z \Big)\\
\sigma (z-\gamma) \left[ \begin{array}{c}
\rho\, v_\theta^{1/2}\\
v_\tau^{1/2}\\
\bzero 
\end{array} \right] 
\end{array} \right]\\
&= \left[ \begin{array}{c}
\sigma^2 \Big( 2\gamma(z-\gamma) + (z-\gamma)^2 - 1 \Big)\\
\sigma (z-\gamma) \left[ \begin{array}{c}
\rho\, v_\theta^{1/2}\\
v_\tau^{1/2}\\
\bzero 
\end{array} \right] 
\end{array} \right].
\end{align*}
It may also be shown, using the definitions of the Hermite polynomials of degrees 1, 2 and 3 (given e.g. by \citeauthor{BNandCox1989}, \citeyear{BNandCox1989}), that 
\begin{align*}
&\int_{0}^{\infty} \int_{- d_m w}^{d_m w} z\, E \big( \widehat{\bm{s}} - \bm{s} \big| \widetilde{\gamma}=z \big) \phi(z-\gamma) dz\, f_W(w)\, dw
\\
&\qquad \qquad \qquad \qquad= \sigma \left[ \begin{array}{c}
\sigma \Big( \gamma\, q_m(\gamma) + k_m(\gamma) + h_m(\gamma) \Big)\\
\rho\, v_\theta^{1/2} q_m(\gamma)\\
v_\tau^{1/2} q_m(\gamma)\\
\bzero  
\end{array} \right],
\end{align*}
where the functions $k_m$, $q_m$ and $h_m$ are defined by (1), (3) and (4), respectively. 
Thus
\begin{equation*}
%\label{cov_eta}
\cov_*(\bm{\eta}) = \sigma^2 \left[ \begin{array}{c}
2\theta - \rho\, \sigma v_\theta^{1/2} \Big( \gamma\, q_m(\gamma) + k_m(\gamma) + h_m(\gamma) \Big)\\
v_\theta \, \big(1 - \rho^2 \, q_m(\gamma)\big) \\
\rho\, v_\theta^{1/2} v_\tau^{1/2} \big( 1-q_m(\gamma) \big) \\
\bzero 
\end{array} \right].
\end{equation*}
The result now follows from \eqref{DeltaMethodSD} and \eqref{V_eta_inverse}.

\hfill $\qed$

\medskip

\noindent \textbf{A2. Proof of Theorem \ref{CPdeltaEfronVarUnknown}}

\smallskip

\noindent Let $G = (\widehat{\theta} - \theta) / (\sigma\, v_\theta^{1/2})$.
The coverage probability of the ${\bf sd}_{\rm delta}$ {\bf interval} is 
\begin{align*}
&P \left( \widetilde{\theta} - t_m(\alpha)\, {\bf sd}_{\rm delta}(\widehat{\gamma}, \widehat{\sigma}) \leq \theta \leq \widetilde{\theta} + t_m(\alpha)\, {\bf sd}_{\rm delta}(\widehat{\gamma}, \widehat{\sigma}) \right) \\
%&= {\rm Pr}\left(  - t_m(\alpha)\, {\bf sd}_{\rm delta}(\widehat{\gamma}, \widehat{\sigma}) \leq \widetilde{\theta} - \theta \leq  t_m(\alpha)\, {\bf sd}_{\rm delta}(\widehat{\gamma}, \widehat{\sigma}) \right) \\
%&= {\rm Pr}\left(  - t_m(\alpha)\, {\bf sd}_{\rm delta}(\widehat{\gamma}, \widehat{\sigma}) \leq \widehat{\theta} - \theta - \rho \, \widehat{\sigma}\, v_\theta^{1/2}\, k_m(\widehat\gamma) \leq  t_m(\alpha)\, {\bf sd}_{\rm delta}(\widehat{\gamma}, \widehat{\sigma}) \right) \\
&= P\left(  - t_m(\alpha)\, \frac{{\bf sd}_{\rm delta}(\widehat{\gamma}, \widehat{\sigma})}{\widehat{\sigma}\, v_\theta^{1/2}} \leq \frac{\widehat{\theta} - \theta}{\widehat{\sigma}\, v_\theta^{1/2}} - \rho \, k_m(\widehat\gamma) \leq  t_m(\alpha)\, \frac{{\bf sd}_{\rm delta}(\widehat{\gamma}, \widehat{\sigma})}{\widehat{\sigma}\, v_\theta^{1/2}} \right) \\
%&= {\rm Pr}\left(  - t_m(\alpha)\, r_{\rm delta}(\widehat{\gamma}) \leq \frac{\widehat{\theta} - \theta}{\sigma\, v_\theta^{1/2}}\, \frac{1}{\widehat{\sigma}/\sigma} - \rho \, k_m(\widehat\gamma) \leq  t_m(\alpha)\,  r_{\rm delta}(\widehat{\gamma}) \right) \\
&= P\left(  - t_m(\alpha)\, r_{\rm delta}(\widehat{\gamma}) \leq \frac{G}{W} - \rho \, k_m(\widehat\gamma) \leq  t_m(\alpha)\,  r_{\rm delta}(\widehat{\gamma}) \right) 
\\
&= P\Bigg(  - t_m(\alpha)\, r_{\rm delta}\left(\frac{\widetilde{\gamma}}{W} \right) \leq \frac{G}{W} - \rho \, k_m\left(\frac{\widetilde{\gamma}}{W} \right) \leq  t_m(\alpha)\,  r_{\rm delta}\left(\frac{\widetilde{\gamma}}{W} \right) \Bigg)
\\
&= \int_{0}^{\infty} \int_{-\infty}^{\infty} P \Bigg(  - t_m(\alpha)\, r_{\rm delta}\left(\frac{\widetilde{\gamma}}{W} \right) \leq \frac{G}{W} - \rho \, k_m\left(\frac{\widetilde{\gamma}}{W} \right) \leq  t_m(\alpha)\,  
r_{\rm delta}\left(\frac{\widetilde{\gamma}}{W} \right) \Bigg|
\\ 
&\qquad \qquad \qquad \qquad \qquad \qquad \qquad \qquad  \qquad  \widetilde{\gamma}=h, W=w \Bigg)  \phi(h-\gamma) \, dh \, f_W(w)\, dw.
\end{align*}
By the substitution theorem for conditional expectations and
since $G$ and $W$ are independent random variables, this is equal to
\begin{align*}
&\int_{0}^{\infty} \int_{-\infty}^{\infty} P\Bigg(  - t_m(\alpha)\, r_{\rm delta}\left(\frac{h}{w} \right) \leq \frac{G}{w} - \rho \, k_m\left(\frac{h}{w}\right) \leq  t_m(\alpha)\,  r_{\rm delta}\left(\frac{h}{w} \right) \Bigg| \widetilde{\gamma}=h \Bigg) \\
&\qquad \qquad \qquad \qquad \qquad \qquad \qquad \qquad \qquad \qquad  \qquad \qquad \qquad \phi(h-\gamma) \, dh \, f_W(w)\, dw.
\end{align*}
Obviously
\begin{align*}
&P\Bigg(  - t_m(\alpha)\, r_{\rm delta}\left(\frac{h}{w} \right) \leq \frac{G}{w} - \rho \, k_m\left(\frac{h}{w} \right) \leq  t_m(\alpha)\,  r_{\rm delta}\left(\frac{h}{w} \right) \Bigg| \widetilde{\gamma}=h \Bigg) 
\\
%&= {\rm Pr}\Bigg(  - t_m(\alpha)\, r_{\rm delta}\left(\frac{h}{w} \right) + \rho \, k_m\left(\frac{h}{w} \right) \leq \frac{G}{w}  \leq  t_m(\alpha)\,  r_{\rm delta}\left(\frac{h}{w} \right) + \rho \, k_m\left(\frac{h}{w} \right) \Bigg| \widetilde{\gamma}=h \Bigg) \\
%&= {\rm Pr}\Bigg(  - w\, t_m(\alpha)\, r_{\rm delta}\left(\frac{h}{w} \right) + w\, \rho \, k_m\left(\frac{h}{w} \right) \leq G  \leq  w\, t_m(\alpha)\,  r_{\rm delta}\left(\frac{h}{w} \right) + w\, \rho \, k_m\left(\frac{h}{w} \right) \Bigg| \widetilde{\gamma}=h \Bigg) \\
&= P \Big( \ell(h, w, \rho) \leq G \leq u(h, w, \rho) \Big| \widetilde{\gamma}=h \Big),
\end{align*}
where the functions $\ell$ and $u$ are defined by \eqref{Def_ell} and \eqref{Def_u}, respectively.
The distribution of $G$, conditional on $\widetilde{\gamma}=h$, is $N\big(\rho(h-\gamma), 1-\rho^2 \big)$. Thus 
$
P \big( \ell(h, w, \rho) \leq G \leq u(h, w, \rho) \, \big| \, \widetilde{\gamma}=h \big) = P \big( \ell(h, w, \rho) \leq \widetilde{G} \leq u(h, w, \rho) \big)$,
where $\widetilde{G} \sim N\big(\rho(h-\gamma), 1-\rho^2 \big)$. 
Therefore the coverage probability $\CP_{\rm delta} (\gamma, \rho)$ is equal to
\begin{align*}
%&= \int_{0}^{\infty} \int_{-\infty}^{\infty} {\rm Pr} \left( \ell(h, w, \rho) \leq \widetilde{G} \leq u(h, w, \rho) \right) \phi(h-\gamma) \, dh \, f_W(w)\, dw \\
\int_{0}^{\infty} \int_{-\infty}^{\infty} \Psi \Big( \ell(h, w, \rho), u(h, w, \rho); \rho(h-\gamma), 1-\rho^2 \Big) \phi(h-\gamma) \, dh \, f_W(w)\, dw.
\end{align*}
The result follows by changing the variable of integration 
of the inner integral to $y = h - \gamma$.

\hfill $\qed$

\medskip

\noindent \textbf{A3. Proof of Theorem \ref{SELdeltaEfronVarUnknown}}

\smallskip

\noindent The scaled expected length $\SEL_{\rm delta}(\gamma, \rho) = E\big(\text{length of } J_{\rm delta}\big) \big/ E\big( \text{length of } I(c_{\rm min}) \big)$.
%\begin{equation*}
%SEL_{\rm delta}(\gamma, \rho) = \frac{E\left(\text{length of } J_{\rm delta}\right)}
%{E\left( \text{length of } I(c_{\rm min}) \right)}.
%\end{equation*}
%%
The length of $J_{\rm delta}$ is $2 \, t_m(\alpha)\, {\bf sd}_{\rm delta}(\widehat{\gamma}, \widehat{\sigma})$. Thus
\begin{align*}
E \big( \text{length of }J_{\rm delta} \big) 
= 2 \, t_m(\alpha) \, \sigma \, v_\theta^{1/2} \, E\left( W\, \, r_{\rm delta}\Big(\frac{\widetilde{\gamma}}{W} \Big)  \right).
\end{align*}
The expected length of $I(c_{\rm min})$ is 
\begin{align*}
E\left( 2\, t_m(1 - c_{\rm min}) \, \widehat{\sigma}\, v_\theta^{1/2} \right) 
%&= 2\, t_m(1 - c_{\rm min}) \, v_\theta^{1/2}\, E\left(\widehat{\sigma}\right)\\
%&= 2\, t_m(1 - c_{\rm min}) \, \sigma \, v_\theta^{1/2}\, E\Big(\frac{\widehat{\sigma}}{\sigma}\Big)\\
= 2\, t_m(1 - c_{\rm min}) \, \sigma \, v_\theta^{1/2}\, E\left(W\right).
\end{align*}
Hence the scaled expected length is
\begin{align*}
{\SEL}_{\rm delta}(\gamma, \rho) 
= \frac{t_m(\alpha) }{t_m(1 - c_{\rm min}) }\, \frac{E\left( W\, \, \displaystyle r_{\rm delta}\Big(\frac{\widetilde{\gamma}}{W} \Big)  \right)}{E\left(W\right)}.
\end{align*}
Since $W$ has the same distribution as $(Q/m)^{1/2}$, where $Q \sim \chi^2_m$,
\begin{align*}
E(W) 
= \left(\frac{m}{2}\right)^{-1/2} \, \frac{\Gamma((m+1)/2)}{\Gamma(m/2)}.
\end{align*}
Hence 
\begin{align*}
&\SEL_{\rm delta}(\gamma, \rho)\\
%&= \frac{t_m(\alpha) }{t_m(1 - c_{\rm min}) }\, \left(\frac{m}{2}\right)^{1/2} \, \frac{\Gamma(m/2)}{\Gamma((m+1)/2)}\, E\left( W\, \, r_{\rm delta}\Big(\frac{\widetilde{\gamma}}{W} \Big) \right) \\
&= \frac{t_m(\alpha) }{t_m(1 - c_{\rm min}) }\, \left(\frac{m}{2}\right)^{1/2} \, \frac{\Gamma(m/2)}{\Gamma((m+1)/2)}\, \int_{0}^{\infty} \int_{-\infty}^{\infty} w\, r_{\rm delta}\left(\frac{h}{w}\right)\, \phi(h-\gamma)\, dh\, f_W(w)\, dw.
\end{align*}
The result follows by changing the variable of integration 
of the inner integral to $y = h - \gamma$.

\hfill $\qed$

\end{document}